# High ductility and transformation-induced-plasticity in stainless steel 304L processed by selective laser melting with low power

E. Polatidis[a*], J. Capek[a], A. Arabi-Hashemi[b], C. Leinenbach[b], M. Strobl[a]

[a]Laboratory for Neutron Scattering and Imaging, Paul Scherrer Institute, Villigen PSI, 5232, Switzerland

[b]Empa, Swiss Federal Laboratories for Materials Science and Technology, Überlandstrasse 129, CH-8600 Dübendorf, Switzerland

* Corresponding author: efthymios.polatidis@psi.ch

**Abstract (100 words)**

304L steel powder was processed by selective laser melting with low power, resulting in nearly random crystallographic texture. In-situ tensile loading and neutron diffraction experiments were undertaken and the results indicate high ductility, despite the presence of porosity, and pronounced strain-induced martensitic transformation. A secondary hardening is observed in the mechanical data due to the martensite carrying significant load, upon its formation. The pronounced martensitic transformation is discussed with respect to the initial and the evolving deformation texture, as revealed by EBSD, and its affinity to ε-martensite formation, under uniaxial loading, which is a precursor for α'-martensite formation.

Selective laser melting (SLM) is an additive layer-by-layer manufacturing process where the component is built by a laser heat-source that melts the powder, which is spread on the base plate by a recoater [1]. This technique is well established for processing complex component geometries and different metallic alloys, amongst which austenitic stainless steels are used extensively. Austenitic stainless steels, such as 316L and 304L exhibit a combination of high strength and ductility and good corrosion resistance. Due to their additional good SLM processability, they are ideal candidates for medical applications profiting e.g. from the possibility of additive manufacturing to produce complex geometries [2,3].

The austenitic steel 304L, due to its relatively low stacking fault energy (SFE), i.e. 18 mJ/m$^2$ as reported in [4], is known to exhibit the transformation induced plasticity (TRIP) effect upon deformation of its wrought form [5–7], where the face-centered-cubic (FCC) parent phase transforms to the hexagonal-closed-packed (hcp) ε-martensite and the cubic-body-centered (BCC) α'-martensite. The TRIP effect has been shown to result in a high rate of macroscopic strain hardening [8,9]. The extend of strain-induced martensite depends on the loading state in combination with the crystallographic texture [10–13]. It was seen that uniaxial loading favors the martensitic transformation in randomly textured, low SFE austenitic steel following the sequence γ →ε → α', where at low strains ε-martensite is the precursor of α'-martensite. The relationship of texture/loading direction is rationalized by comparing the Schmid factor (SF) of the leading partial dislocation to the SF of the trailing partial dislocation. For crystallographic orientations for which the SF of the leading partial dislocation is higher than the SF of the trailing, extensive stacking faults form allowing the formation of ε-martensite [10,11].

The SLM processing parameters can largely influence the microstructure, porosity, surface roughness and eventually the mechanical properties of the components. The crystallographic texture has been also shown to be sensitive to the SLM processing parameters and can vary from random using low energy input [14–16] to strong texture [14,16] during SLM processing. Extensive studies have been carried out on the optimization of the SLM process parameters for the 316L steel [17,14,18–25], with respect to porosity minimization, mechanical properties optimization and occurrence of residual stress. However, not as many studies exist for SLM-processing of the 304L steel [26–29].

In a recent study, a good combination of strength and ductility was obtained in the as-SLM processed material due to the optimization of the SLM process parameters which resulted in the presence of cellular structures, small grain size and martensite, all of which are attributed to the fast cooling rates [29]. However, the deformation mechanisms, which gave rise to the combination of high strength and good ductility were not investigated in depth. In more recent studies on 304L processed by direct energy deposition (DED), it was shown that the transformation is suppressed, due to high nitrogen content of the powder used for additive manufacturing processes [30,31]. Hence, the dominant deformation behavior of materials processed with some additive manufacturing methods can be different to that of wrought materials.

In the present work, the SLM processing parameters are selected such that the as-SLM processed 304L steel exhibits nearly random crystallographic texture. Based on previous observation this crystallographic texture is favorable for exhibiting pronounced TRIP effect under uniaxial tension, for low SFE wrought austenitic steels [11]. The SLM-processed material is investigated using in-situ uniaxial deformation with neutron diffraction to reveal the deformation mechanisms of the SLM-processed material. The mechanical properties are discussed in relation to the residual stress that is present in the SLM-processed material, the evolving deformation texture and the TRIP effect.

For the SLM process, gas-atomized 304L powder with maximum particle size of 45 μm, was purchased from Goodfellow, UK. The SLM-fabrication was performed using a Sisma MySint 100 with a spot size of 55 μm, laser power 175 W and laser scan velocity of 1200 mm/s. A "chess-board" scanning pattern (squares of 4×4mm$^2$) for each layer was rotated 90 degrees and shifted 1 mm in x and y with respect to the previous layer. The obtained density is ~98%, as characterized by optical microscopy on sample cubes produced with the same conditions (cf. Fig. S1 of the supplementary materials). Cylinders of 14 mm in diameter and 86 mm in height were built with their long direction parallel to the building direction. "Dogbone" specimens for uniaxial tension tests were then machined from the as-built cylinders, the geometry is shown in Fig. S2 of the supplementary material.

Residual stress characterization by neutron diffraction of the SLM processed material was undertaken on the POLDI instrument at SINQ at the Paul Scherrer Institute, Switzerland, using a 3.8×3.8×3.8 mm³ gauge volume. The residual strain was measured along the radial and axial direction along a line in the center of the cylinder scanned with 5 steps of 12 mm step size over +/- 24 mm from the center of the as-built cylinder and only at the central point for the machined dogbone specimen (shown in Fig. S3 of the supplementary material). For the residual stress characterization, the lattice strain was calculated using the {311} lattice plane family as it best represents the bulk elastic properties for FCC materials [32,33]. Reference measurements were undertaken on an annealed cylinder at 450°C for 5 hours for obtaining the strain free interplanar lattice spacing, $d^0_{\{311\}}$. Based on the fact that the interplanar lattice spacing of the reference measurement shows values very close to the ones obtained from a measurement on the as-received gas-atomized powder, it can be considered "strain-free". The value of $d^0_{\{311\}}$ did not vary significantly from the top to the bottom of the cylinder, implying no considerable chemical variations, and therefore an average value was used for calculating the residual strain for all measurement points. The residual elastic strain, $\varepsilon_{\{311\}}$, was then calculated from the interplanar lattice spacing, $d_{\{311\}}$, of each point along an SLM-built cylinder, as follows:

$$\varepsilon_{\{311\}} = \frac{d_{\{311\}} - d^0_{\{311\}}}{d^0_{\{311\}}} \qquad (1)$$

Residual stress is then calculated using Hooke's law. For the diffraction elastic constants, the lattice strain evolution of the {311} lattice plane family obtained from the (longitudinal) in situ deformation test was fitted in the elastic regime by a line, the slope of which is equal to the Young's modulus, i.e. $E_{\{311\}}$=175 GPa (cf. Fig. S4 of the supplementary materials). The ratio of the lattice strain in the elastic regime between the transverse and longitudinal measurements was used to calculate the Poisson's ratio of the {311} lattice plane family, i.e. $\nu_{\{311\}}$=0.3 as show in Fig. S4 of the supplementary material.

It is seen that the as-SLM-processed material exhibits high compressive stress along the building direction, which is in good agreement with the reports in literature [34]. The machining process of the dogbone sample relieves approximately 80% of the residual stress, as shown in Fig. S3 of the supplementary material.

For the in situ uniaxial deformation and neutron diffraction experiments, the dogbone specimens were deformed with 0.01 mm/min displacement rate. The neutron diffraction measurements were undertaken upon stopping and holding the displacement at pre-defined force values (in the elastic regime) and pre-defined strain values (in the plastic regime). The in-situ diffraction neutron measurements were only undertaken at the center of the dogbone sample. For both residual stress and in situ deformation tests, the neutron data were analyzed and fitted using Mantid [35]. The data were analyzed qualitatively in terms of appearance of new reflections due to the presence of martensite. The evolution of the elastic lattice strain $\varepsilon_{hkl}$ was also determined by the relative change of the interplanar lattice spacing $d_{hkl}$ with respect to $d_{0hkl}$ which is the initial value prior to deformation:

$$\varepsilon_{hkl} = \frac{d_{hkl} - d_{0hkl}}{d_{0hkl}} \qquad (2)$$

Using $d_{0hkl}$ which is the initial value for austenite, prior to deformation, does not account for the lattice strain, already present due to the low residual stress in the material. However, it does not affect assessing the slope in the elastic regime for calculating the diffraction elastic constant, or for assessing which phase accumulates more strain once martensite forms. The error of $\varepsilon_{hkl}$ is calculated from the error (fitting uncertainty) of $d_{0hkl}$ and $d_{hkl}$.

The lattice strain evolution in martensite (shown in Fig. 3-b) is calculated beyond 0.23 true strain that it first appears and hence the reference interplanar lattice spacing is taken as the d-spacing value at 0.23 true strain, i.e. $d_{0.23hkl}$. In addition, the lattice strain evolution of austenite beyond the appearance of martensite (also shown in Fig. 3-b) is calculated in the same way as for martensite:

$$\varepsilon_{hkl} = \frac{d_{hkl} - d_{0.23hkl}}{d_{0.23hkl}} \qquad (3)$$

The error of $\varepsilon_{hkl}$ is calculated from the error (fitting uncertainty) of $d_{0.23hkl}$ and $d_{hkl}$.

The as-SLM-processed and the deformed materials were further characterized by electron backscattered diffraction (EBSD). Samples were cut from the top (away from the base plate), bottom (close to the base plate) and the middle of the deformed dogbone sample. The samples were ground with 1200 grit SiC paper and then electropolished for 12 s with a 16:3:1 (by volume) ethanol, glycerol and perchloric acid solution at 42 V. A field emission gun scanning electron microscope (FEG SEM) Zeiss ULTRA 55 equipped with EDAX Hikari Camera operated at 20 kV in high current mode with 120 µm aperture was used. The EBSD raw data was post-processed using the EDAX OIM Analysis 7.3 software.

The as-SLM-processed material is fully austenitic, as seen in the inverse pole figure (IPF) map shown in Fig. 1 and in the neutron diffraction pattern shown in Fig. 2-b, which is in contrast with previous works where the as-SLM-processed material exhibited a small fraction of martensite, the presence of which was attributed to the fast cooling rates [29,36]. Fig. 1 shows the microstructure of the as-SLM-processed material, parallel to the building direction and the inverse pole figures (IPFs) parallel to the three principal directions (i.e. building direction-BD, transverse direction-TD and normal to the sample surface-ND). The crystallographic texture is seen to be nearly random, as previously seen for low energy SLM-processed stainless steels [14,17] and when the laser scanning pattern of each layer is rotated 90 degrees with respect to the previous layer [37].

Fig. 2-a shows the true stress versus true strain mechanical data and the work hardening rate obtained from a continuous (uninterrupted) test performed ex-situ. The mechanical data from the in-situ test coincide very well with the data obtained from the ex-situ test. The yield stress, $YS_{0.2}$ is 440 MPa which is slightly lower than for the SLM-processed 304L alloy reported in [27]. However, such variations can be due to different magnitude of residual stress and/or microstructural differences (e.g. grain size, grain morphology, crystallographic texture etc.). Upon deformation, a secondary strain hardening

regime is observed after approximately 0.23 true strain as shown in Fig. 2-a. The strengthening is also apparent in the work hardening rate also plotted in Fig. 2-a. This hardening coincides with the onset of the martensitic transformation, as apparent by the appearance of the (110) reflection of α'-martensite shown in Fig. 2-b. Two very weak reflections, i.e. (10$\bar{1}$0) at 2.9 Å and (10$\bar{1}$1) at 3.2 Å, corresponding to ε-martensite also appear at true strain higher than 0.23. The latter observations indicate the formation ε-martensite, additional to α'-martensite, as previously seen for low SFE austenitic steels [11,38,39]. During the initial stages of plastic deformation, diffraction peak broadening is observed as indicative of plasticity by dislocation density increase, slip, or stacking fault formation as seen in Fig. S5 of the supplementary material. Significant amount of deformation induced α'-martensite is observed in the SLM-processed 304L material in contrast to previous observations on 304L processed by DED, where strain-induced martensite was suppressed [30,31].

Fig. 3-a shows the evolution of the lattice strain of austenite with increasing strain. Until the macroscopic yield point, $YS_{0.2}$, a typical mechanical behavior for polycrystalline FCC materials is seen, where the {111} family of planes is the stiffest and the {200} family is the most compliant and the {311} family exhibits an intermediate mechanical behavior. The secondary hardening can be explained by comparing the austenite and martensite lattice strain beyond 0.23 true strain. As shown in Fig.3-b, the martensite lattice planes accumulate strain at higher rate than austenite. For instance, by comparing the lattice strain in the $\{311\}_{FCC}$ and $\{211\}_{BCC}$ lattice plane families, which are typically considered as representative of the bulk material properties, it is apparent that martensite carries more load than austenite.

It is known that the crystallographic orientation changes by deformation and a strong <111>-texture evolves parallel to the loading direction in uniaxial tensile deformation [11]; the evolved texture at 0.26 and 0.42 true strain is shown in Fig. S6 of the supplementary material. The low magnification EBSD maps in Fig 4-a and 4-b show the austenite grains which have crystallographic orientations favorable for the splitting of the partial dislocations in blue (i.e. the orientations of grains, in the direction parallel the loading direction, for which the SF of the leading partial dislocations is higher than the SF of the trailing partial dislocations) and the austenite grains which have crystallographic orientations not favorable for the splitting of the partial dislocations (i.e. the orientations of grains, in the direction parallel the loading direction, for which the SF of the leading partial dislocations is lower than the SF of the trailing partial dislocations) in red. The α'-martensite grains are shown with yellow. Figures 4a and -b confirm that the majority of the austenite grains (at 0.26 and 0.42 true strain) have crystallographic orientations that favor the formation of stacking faults and ε-martensite (shown with blue color in Fig. 4-b, -c and -d). A significant amount of α'-martensite is observed at 0.26 true strain and even more is formed at 0.42 true strain (Fig. 4-d), which is in good agreement with the neutron diffraction results. The presence of ε-martensite is evidenced (with high confidence index-CI) in the high resolution EBSD map at high magnification presented in Fig. 4-c. It shows an austenite grain with favorable crystallographic orientation for the formation of ε-martensite. ε-martensite plates shown with green color form on the $(\bar{1}11)$ slip plane of this grain. Furthermore, α'-martensite is evidenced (with high CI) inside the bands of ε-martensite. According to Tian et al. [38] α'-martensite can form in individual ε-martensite in very low SFE steels, indicative of the instability of austenite. It is generally reported that ε-martensite can be a transient phase to α'-martensite [11,40,41] but it is not a necessary precursor, as α'-martensite can appear at intercepts of shear bands in the form of stacking faults or twinning [13,39,42]. The perquisites of all these nucleation mechanisms of α'-martensite is the splitting of the partial dislocations. Besides shear bands formed by the splitting of partial dislocations, α'-martensite is often seen at grain boundaries [39], as also evidenced in Fig. 4-c at the bottom of the grain.

In conclusion, good ductility, despite the observed porosity, and pronounced martensite formation was observed under uniaxial deformation of SLM-processed 304L material produced with low laser power and chessboard laser scanning pattern. The crystallographic texture of the as-SLM-processed material is nearly random while upon deformation a strong <111>-texture evolves with increasing deformation. A majority of grains are favorably oriented so that the leading partial dislocation experiencing higher stress than the trailing partial dislocation, allowing the formation of extended stacking faults and ε-martensite which facilitate the formation of α'-martensite, while α'-martensite can form in single bands of ε-martensite or at grain boundaries. The appearance of strain-induced martensite results in a secondary work-hardening regime as martensite carries more load than austenite.

**Figures and captions**

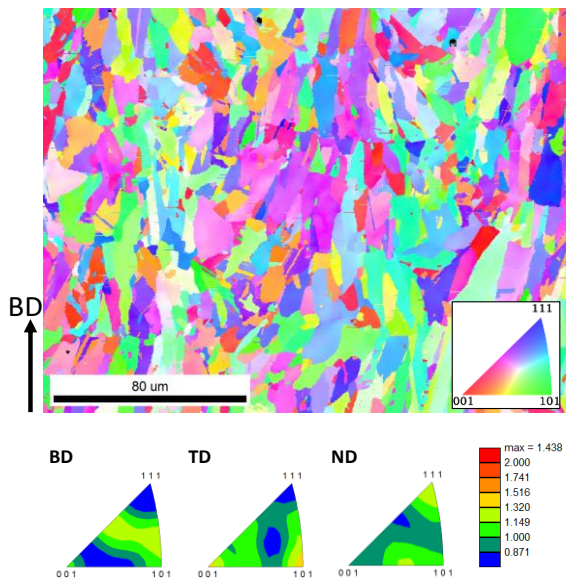

Fig. 1 IPF map of the austenite phase of the as-SLM-processed 304L steel in the direction parallel to the building direction (BD), obtained at the center of the cylinder. The inverse pole figures show the crystallographic orientation with respect to BD, transverse direction (TD) and normal direction (ND).

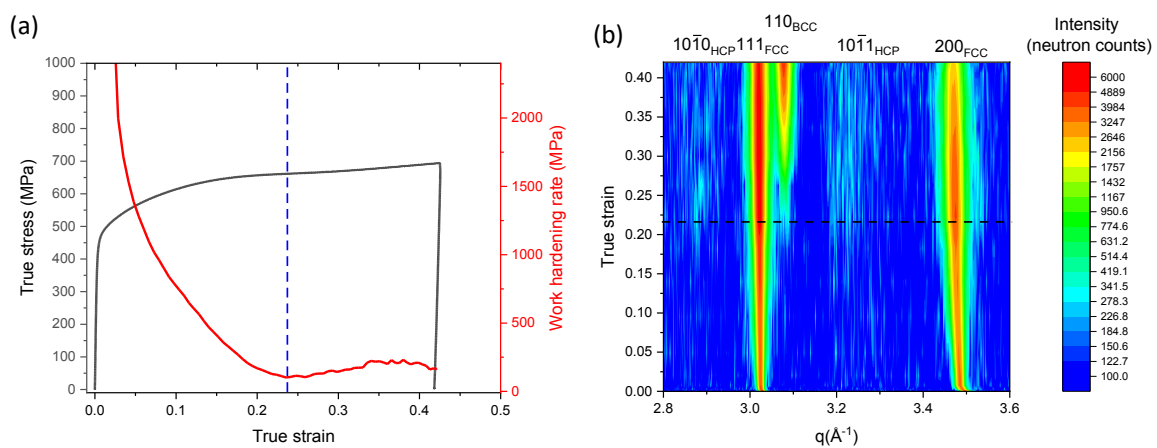

Fig. 2 a) True stress-strain and work hardening rate of an ex-situ continuous test showing hardening after approximately 0.23 true strain (indicated with a blue dashed line). (b) Evolution of neutron diffraction patterns showing the martensite formation (appearance of the $110_{BCC}$, $1010_{HCP}$ and $1011_{HCP}$ reflections) after approximately 0.23 true strain (indicated with a black dashed line).

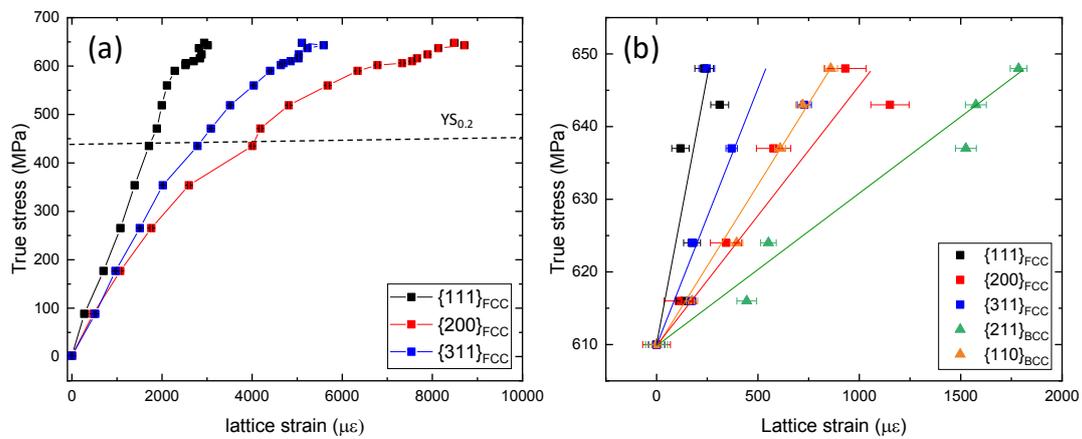

Fig. 3 (a) Lattice strain evolution in austenite. (b) Lattice strain evolution from 0.23 true strain until 0.42 true strain showing that martensite accumulates lattice strain at higher rate than austenite. The lines in (b) are liner fits to the experimental data for guiding the eye.

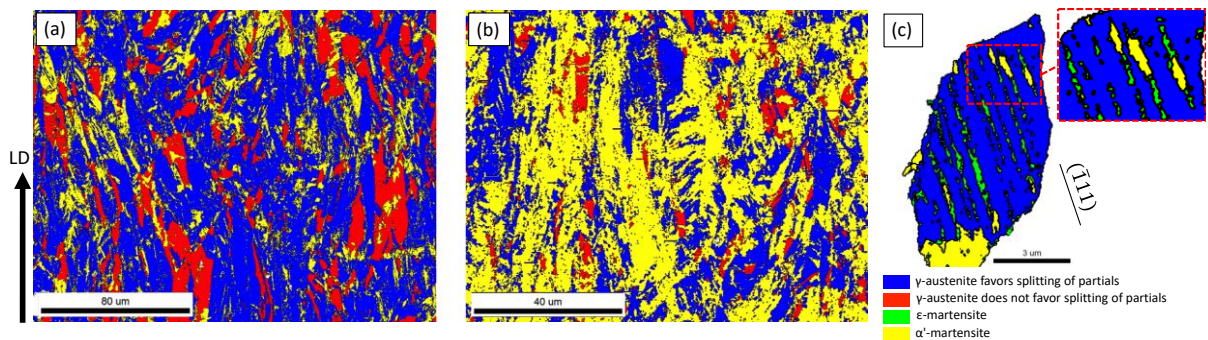

Fig. 4 EBSD maps showing with blue the crystallographic orientations, in the direction parallel to the building direction (parallel to the loading direction-LD), that favor the formation of stacking faults and ε-martensite (i.e. the SF of the leading partial dislocations is higher than the SF of the trailing partial dislocations) under uniaxial tension along the given LD at a) 0.26 true strain and b) 0.42 true strain. The orientations that do not favor the formation of stacking faults and ε-martensite are given in red, α'-martensite is given in yellow and ε-martensite in green. c) Detail of a grain which is favorably orientated for the formation of ε-martensite, at 0.26 true strain, showing ε-martensite plates being formed along the $(\bar{1}11)$ slip plane traces, while α'-martensite forms within single bands of ε-martensite.

**Acknowledgements**

JC thanks the financial support from the Strategic Focus Area Advanced Manufacturing (SFA-AM) initiative of the ETH Board. This project has received funding from the European Union's Horizon 2020 research and innovation programme under the Marie Skłodowska-Curie grant agreement No 701647.

**References**

[1] D. Herzog, V. Seyda, E. Wycisk, C. Emmelmann, Additive manufacturing of metals, Acta Mater. 117 (2016) 371–392. doi:10.1016/j.actamat.2016.07.019.

[2] F. Bartolomeu, M. Buciumeanu, E. Pinto, N. Alves, O. Carvalho, F.S. Silva, G. Miranda, 316L stainless steel mechanical and tribological behavior—A comparison between selective laser melting, hot pressing and conventional casting, Addit Manuf. 16 (2017) 81–89. doi:10.1016/j.addma.2017.05.007.

[3] D. Kong, X. Ni, C. Dong, X. Lei, L. Zhang, C. Man, J. Yao, X. Cheng, X. Li, Bio-functional and anti-corrosive 3D printing 316L stainless steel fabricated by selective laser melting, Mater Des. 152 (2018) 88–101. doi:10.1016/j.matdes.2018.04.058.

[4] R.E. Schramm, R.P. Reed, Stacking fault energies of seven commercial austenitic stainless steels, Metall Mater Trans A. 6 (1975) 1345. doi:10.1007/BF02641927.

[5] G.B. Olson, M. Cohen, Kinetics of strain-induced martensitic nucleation, Metall Mater Trans A. 6 (1975) 791. doi:10.1007/BF02672301.

[6] F. Lecroisey, A. Pineau, Martensitic transformations induced by plastic deformation in the Fe-Ni-Cr-C system, Metall Mater Trans B. 3 (1972) 391–400. doi:10.1007/BF02642042.

[7] A.K. De, D.C. Murdock, M.C. Mataya, J.G. Speer, D.K. Matlock, Quantitative measurement of deformation-induced martensite in 304 stainless steel by X-ray diffraction, Scripta Mater. 50 (2004) 1445–1449. doi:10.1016/j.scriptamat.2004.03.011.

[8] J.B. Leblond, Mathematical modelling of transformation plasticity in steels II: Coupling with strain hardening phenomena, Int J Plast. 5 (1989) 573–591. doi:10.1016/0749-6419(89)90002-8.

[9] R.G. Stringfellow, D.M. Parks, A self-consistent model of isotropic viscoplastic behavior in multiphase materials, Int J Plast. 7 (1991) 529–547. doi:10.1016/0749-6419(91)90043-X.

[10] S. Martin, C. Ullrich, D. Rafaja, Deformation of Austenitic CrMnNi TRIP/TWIP Steels: Nature and Role of the ε−martensite, Mater Today: Proceedings. 2 (2015) S643–S646. doi:10.1016/j.matpr.2015.07.366.

[11] E. Polatidis, W.-N. Hsu, M. Šmíd, T. Panzner, S. Chakrabarty, P. Pant, H. Van Swygenhoven, Suppressed martensitic transformation under biaxial loading in low stacking fault energy metastable austenitic steels, Scripta Mater. 147 (2018) 27–32. doi:10.1016/j.scriptamat.2017.12.026.

[12] M. Zecevic, M.V. Upadhyay, E. Polatidis, T. Panzner, H. Van Swygenhoven, M. Knezevic, A crystallographic extension to the Olson-Cohen model for predicting strain path dependence of martensitic transformation, Acta Mater. (2019). doi:10.1016/j.actamat.2018.12.060.

[13] E. Polatidis, M. Šmíd, W.-N. Hsu, M. Kubenova, J. Capek, T. Panzner, H. Van Swygenhoven, The interplay between deformation mechanisms in austenitic 304 steel during uniaxial and equibiaxial loading, Materials Science and Engineering: A. 764 (2019) 138222. doi:10.1016/j.msea.2019.138222.

[14] T. Niendorf, S. Leuders, A. Riemer, H.A. Richard, T. Tröster, D. Schwarze, Highly Anisotropic Steel Processed by Selective Laser Melting, Metall Mater Trans B. 44 (2013) 794–796. doi:10.1007/s11663-013-9875-z.

[15] Y.M. Wang, T. Voisin, J.T. McKeown, J. Ye, N.P. Calta, Z. Li, Z. Zeng, Y. Zhang, W. Chen, T.T. Roehling, R.T. Ott, M.K. Santala, P.J. Depond, M.J. Matthews, A.V. Hamza, T. Zhu, Additively

manufactured hierarchical stainless steels with high strength and ductility, Nat Mater. 17 (2018) 63–71. doi:10.1038/nmat5021.

[16] Z. Sun, X. Tan, S.B. Tor, C.K. Chua, Simultaneously enhanced strength and ductility for 3D-printed stainless steel 316L by selective laser melting, NPG Asia Mater. 10 (2018) 127–136. doi:10.1038/s41427-018-0018-5.

[17] T. Niendorf, F. Brenne, Steel showing twinning-induced plasticity processed by selective laser melting — An additively manufactured high performance material, Mater Charact. 85 (2013) 57–63. doi:10.1016/j.matchar.2013.08.010.

[18] B. Zhang, L. Dembinski, C. Coddet, The study of the laser parameters and environment variables effect on mechanical properties of high compact parts elaborated by selective laser melting 316L powder, Mater Sci Eng A. 584 (2013) 21–31. doi:10.1016/j.msea.2013.06.055.

[19] A.S. Wu, D.W. Brown, M. Kumar, G.F. Gallegos, W.E. King, An Experimental Investigation into Additive Manufacturing-Induced Residual Stresses in 316L Stainless Steel, Metall Mater Trans A. 45 (2014) 6260–6270. doi:10.1007/s11661-014-2549-x.

[20] J.A. Cherry, H.M. Davies, S. Mehmood, N.P. Lavery, S.G.R. Brown, J. Sienz, Investigation into the effect of process parameters on microstructural and physical properties of 316L stainless steel parts by selective laser melting, Int J Adv Manuf Technol. 76 (2015) 869–879. doi:10.1007/s00170-014-6297-2.

[21] Z. Sun, X. Tan, S.B. Tor, W.Y. Yeong, Selective laser melting of stainless steel 316L with low porosity and high build rates, Mater Des. 104 (2016) 197–204. doi:10.1016/j.matdes.2016.05.035.

[22] R. Casati, J. Lemke, M. Vedani, Microstructure and Fracture Behavior of 316L Austenitic Stainless Steel Produced by Selective Laser Melting, J Mater Sci Technol. 32 (2016) 738–744. doi:10.1016/j.jmst.2016.06.016.

[23] E. Liverani, S. Toschi, L. Ceschini, A. Fortunato, Effect of selective laser melting (SLM) process parameters on microstructure and mechanical properties of 316L austenitic stainless steel, J Mater Process Tech. 249 (2017) 255–263. doi:10.1016/j.jmatprotec.2017.05.042.

[24] T. Simson, A. Emmel, A. Dwars, J. Böhm, Residual stress measurements on AISI 316L samples manufactured by selective laser melting, Addit Manuf. 17 (2017) 183–189. doi:10.1016/j.addma.2017.07.007.

[25] C. Qiu, M.A. Kindi, A.S. Aladawi, I.A. Hatmi, A comprehensive study on microstructure and tensile behaviour of a selectively laser melted stainless steel, Sci Rep. 8 (2018) 7785. doi:10.1038/s41598-018-26136-7.

[26] D. l. Bourell, K. Abd-Elghany, Property evaluation of 304L stainless steel fabricated by selective laser melting, Rapid Prototyp J. 18 (2012) 420–428. doi:10.1108/13552541211250418.

[27] K. Guan, Z. Wang, M. Gao, X. Li, X. Zeng, Effects of processing parameters on tensile properties of selective laser melted 304 stainless steel, Mater Des. 50 (2013) 581–586. doi:10.1016/j.matdes.2013.03.056.

[28] D.W. Brown, D.P. Adams, L. Balogh, J.S. Carpenter, B. Clausen, G. King, B. Reedlunn, T.A. Palmer, M.C. Maguire, S.C. Vogel, In Situ Neutron Diffraction Study of the Influence of Microstructure on the Mechanical Response of Additively Manufactured 304L Stainless Steel, Metall Mater Trans A. 48 (2017) 6055–6069. doi:10.1007/s11661-017-4330-4.

[29] Q.B. Nguyen, Z. Zhu, F.L. Ng, B.W. Chua, S.M.L. Nai, J. Wei, High mechanical strengths and ductility of stainless steel 304L fabricated using selective laser melting, J Mater Sci Technol. 35 (2019) 388–394. doi:10.1016/j.jmst.2018.10.013.

[30] Z. Wang, T.A. Palmer, A.M. Beese, Effect of processing parameters on microstructure and tensile properties of austenitic stainless steel 304L made by directed energy deposition additive manufacturing, Acta Mater. 110 (2016) 226–235. doi:10.1016/j.actamat.2016.03.019.

[31] Z. Wang, A.M. Beese, Stress state-dependent mechanics of additively manufactured 304L stainless steel: Part 1 – characterization and modeling of the effect of stress state and texture on

microstructural evolution, Mater Sci Eng A. 743 (2019) 811–823. doi:10.1016/j.msea.2018.11.094.

[32] B. Clausen, T. Lorentzen, T. Leffers, Self-consistent modelling of the plastic deformation of f.c.c. polycrystals and its implications for diffraction measurements of internal stresses, Acta Mater. 46 (1998) 3087–3098. doi:10.1016/S1359-6454(98)00014-7.

[33] T. Pirling, G. Bruno, P.J. Withers, SALSA—A new instrument for strain imaging in engineering materials and components, Mater Sci Eng A. 437 (2006) 139–144. doi:10.1016/j.msea.2006.04.083.

[34] K. An, L. Yuan, L. Dial, I. Spinelli, A.D. Stoica, Y. Gao, Neutron residual stress measurement and numerical modeling in a curved thin-walled structure by laser powder bed fusion additive manufacturing, Mater Des. 135 (2017) 122–132. doi:10.1016/j.matdes.2017.09.018.

[35] O. Arnold, J.C. Bilheux, J.M. Borreguero, A. Buts, S.I. Campbell, L. Chapon, M. Doucet, N. Draper, R. Ferraz Leal, M.A. Gigg, V.E. Lynch, A. Markvardsen, D.J. Mikkelson, R.L. Mikkelson, R. Miller, K. Palmen, P. Parker, G. Passos, T.G. Perring, P.F. Peterson, S. Ren, M.A. Reuter, A.T. Savici, J.W. Taylor, R.J. Taylor, R. Tolchenov, W. Zhou, J. Zikovsky, Mantid—Data analysis and visualization package for neutron scattering and µ SR experiments, Nucl Instrum Methods Phys Res A. 764 (2014) 156–166. doi:10.1016/j.nima.2014.07.029.

[36] C. Haase, J. Bültmann, J. Hof, S. Ziegler, S. Bremen, C. Hinke, A. Schwedt, U. Prahl, W. Bleck, Exploiting Process-Related Advantages of Selective Laser Melting for the Production of High-Manganese Steel, Materials. 10 (2017) 56. doi:10.3390/ma10010056.

[37] L. Thijs, K. Kempen, J.-P. Kruth, J. Van Humbeeck, Fine-structured aluminium products with controllable texture by selective laser melting of pre-alloyed AlSi10Mg powder, Acta Mater. 61 (2013) 1809–1819. doi:10.1016/j.actamat.2012.11.052.

[38] Y. Tian, O.I. Gorbatov, A. Borgenstam, A.V. Ruban, P. Hedström, Deformation Microstructure and Deformation-Induced Martensite in Austenitic Fe-Cr-Ni Alloys Depending on Stacking Fault Energy, Metall Mater Trans A. 48 (2017) 1–7. doi:10.1007/s11661-016-3839-2.

[39] A. Das, S. Sivaprasad, M. Ghosh, P.C. Chakraborti, S. Tarafder, Morphologies and characteristics of deformation induced martensite during tensile deformation of 304 LN stainless steel, Mater Sci Eng A. 486 (2008) 283–286. doi:10.1016/j.msea.2007.09.005.

[40] M. Humbert, B. Petit, B. Bolle, N. Gey, Analysis of the γ–ε–α' variant selection induced by 10% plastic deformation in 304 stainless steel at −60°C, Materials Science and Engineering: A. 454–455 (2007) 508–517. doi:10.1016/j.msea.2006.11.112.

[41] T.-H. Lee, H.-Y. Ha, J.-Y. Kang, J. Moon, C.-H. Lee, S.-J. Park, An intersecting-shear model for strain-induced martensitic transformation, Acta Materialia. 61 (2013) 7399–7410. doi:10.1016/j.actamat.2013.08.046.

[42] J.-Y. Choi, W. Jin, Strain induced martensite formation and its effect on strain hardening behavior in the cold drawn 304 austenitic stainless steels, Scripta Materialia. 36 (1997) 99–104. doi:10.1016/S1359-6462(96)00338-7.

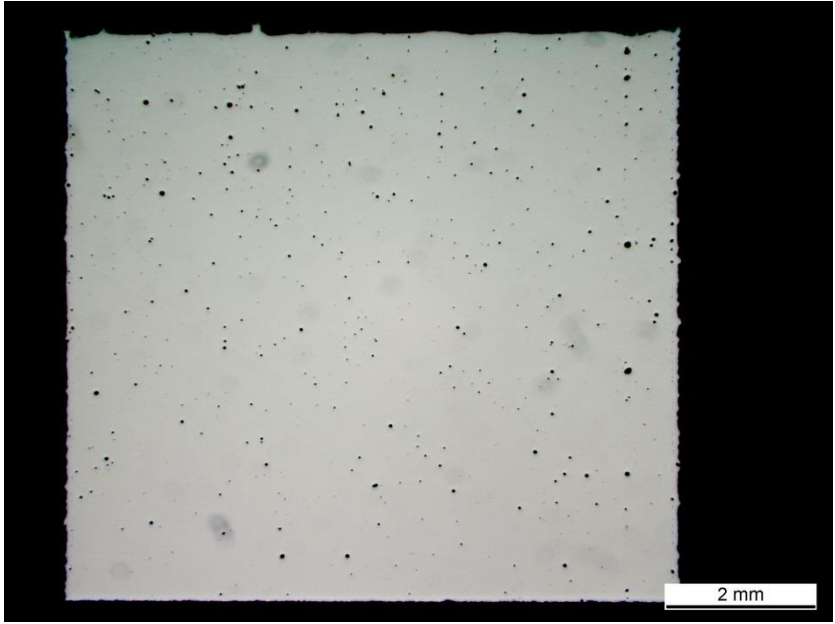

S1. Optical micrograph showing the porosity of a test cube built with the same processing parameters as for the cylindrical samples.

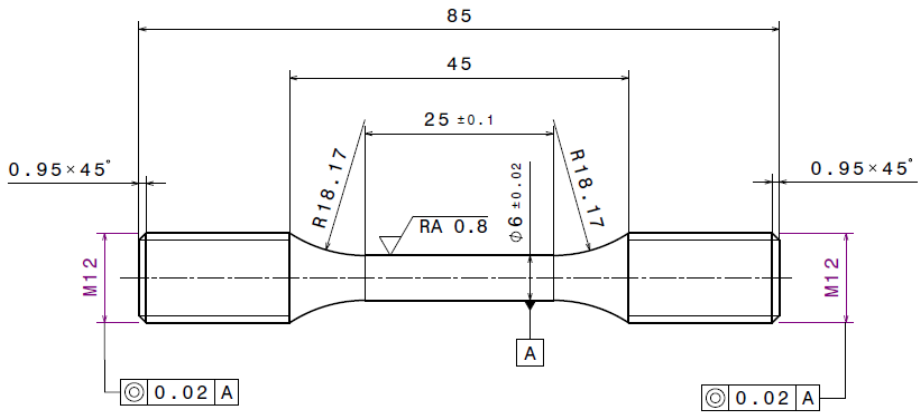

S2. The cylindrical dogbone sample geometry.

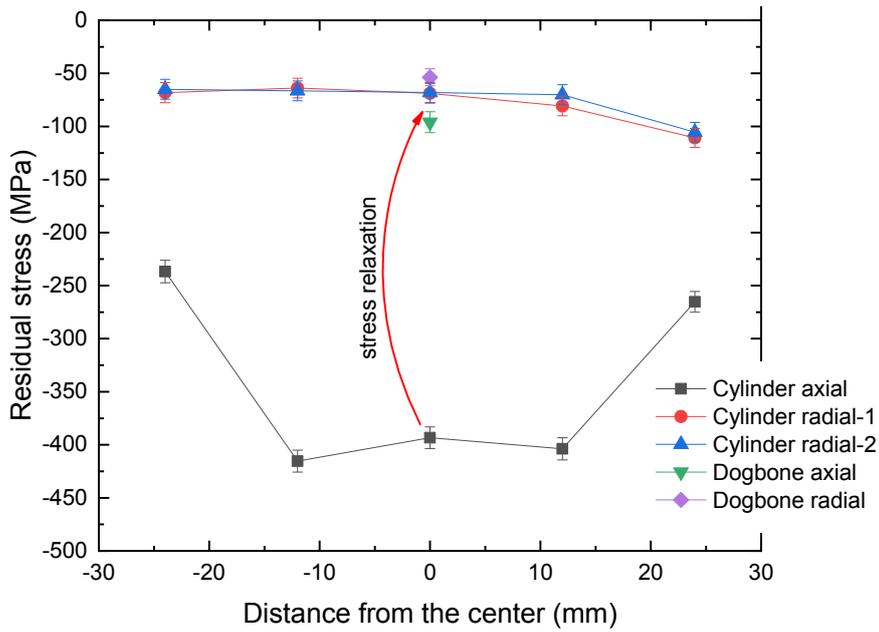

S3. Residual stress profile in the as-built cylinder, along a line at the center of the cylinder with a step size of 12 mm and at the center of the cylindrical dogbone sample. Strong axial (parallel to the building direction) compressive stress is observed in the as-SLM processed material. The residual stress relaxes by machining the dogbone specimen.

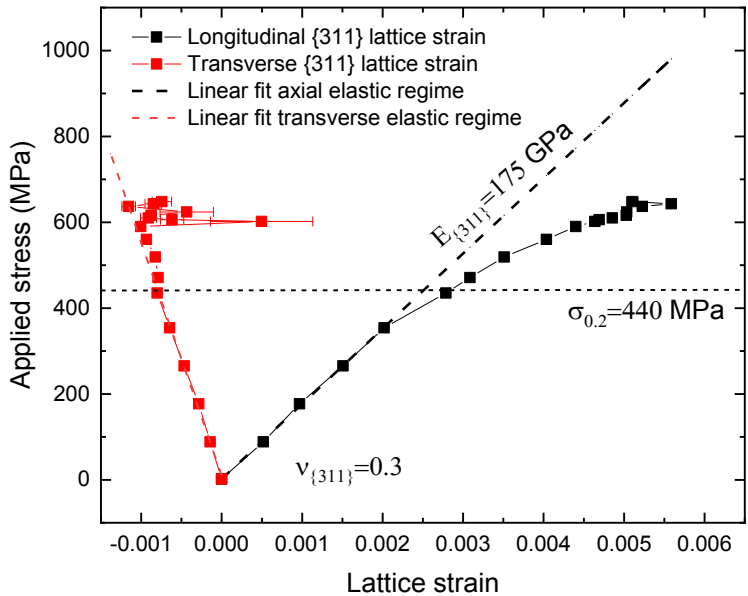

S4. Lattice strain evolution with applied stress for longitudinal (black) and transverse directions (red). Linear fits are applied through the first 5 data points depicting the elastic modulus of the {311} lattice plane family. The elastic strain ratio (i.e. Poisson's ratio) of the transverse over the longitudinal directions equals 0.3.

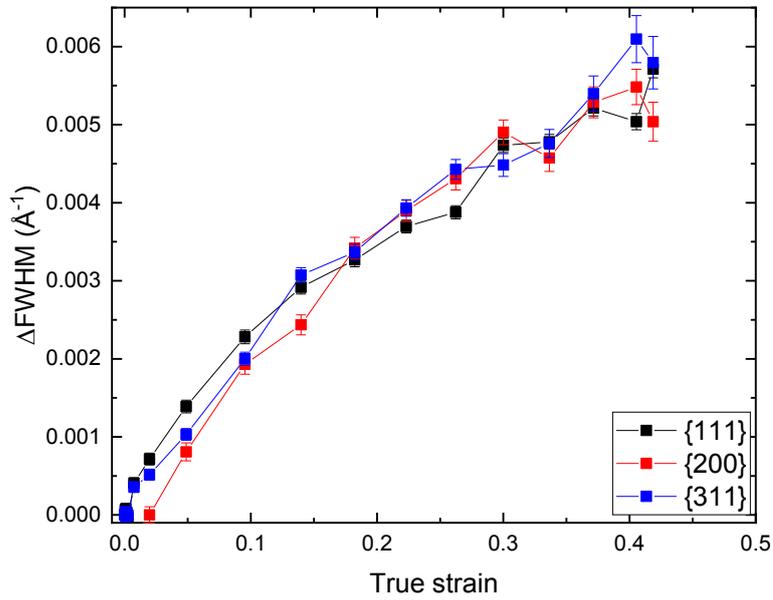

S5. Evolution of peak broadening for various grain families as a function of true strain.

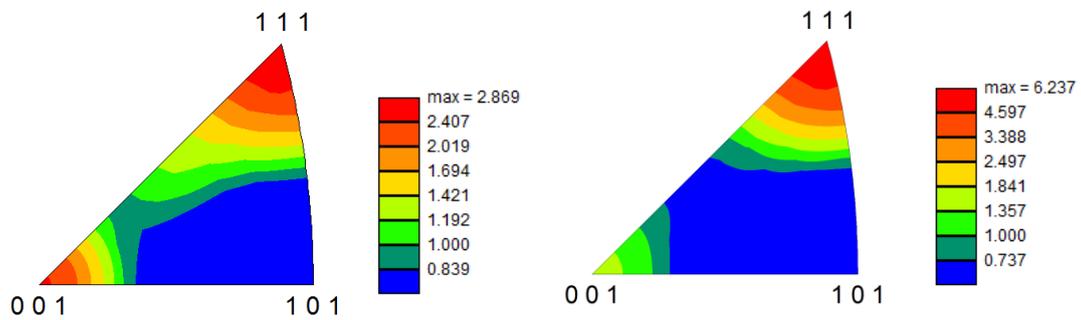

S6. IPFs of the austenite phase of deformed material in the direction parallel to the loading direction showing the increasingly strong <111> -texture with increasing true strain.